\documentclass[aps,prl,reprint,showpacs,lengthcheck,longbibliography,superscriptaddress]{revtex4-1}
\usepackage[T1]{fontenc}

\usepackage{graphicx}
\usepackage{float}
\usepackage{amsmath}
\usepackage{amssymb}
\usepackage[usenames,dvipsnames]{xcolor}
\usepackage[colorlinks,linkcolor=blue,citecolor=blue,urlcolor=blue]{hyperref}
\usepackage{microtype}

\def\equationautorefname~#1\null{Equation~(#1)\null}

\begin{document}

\title{Light-Forbidden Transitions in Plasmon-Emitter Coupling}
\author{A. Cuartero-Gonz\'alez}
\affiliation{Departamento de F\'isica Te\'orica de la Materia
Condensada and Condensed Matter Physics Center (IFIMAC),
Universidad Aut\'onoma de Madrid, E- 28049 Madrid, Spain}
\author{A. I. Fern\'andez-Dom\'inguez}\email{a.fernandez-dominguez@uam.es}
\affiliation{Departamento de F\'isica Te\'orica de la Materia
Condensada and Condensed Matter Physics Center (IFIMAC),
Universidad Aut\'onoma de Madrid, E- 28049 Madrid, Spain}

\begin{abstract}
We investigate the impact that light-forbidden exciton transitions
have in the near-field population dynamics and far-field
scattering spectrum of hybrid plasmon-emitter systems.
Specifically, we consider a V-type quantum emitter, sustaining one
dipolar and one quadrupolar (dipole-inactive) excited states,
placed at the nanometric gap of a particle-on-a-mirror metallic
cavity. Our fully analytical description of plasmon-exciton
coupling for both exciton transitions enables us to reveal the
conditions in which the presence of the latter greatly alters the
Purcell enhancement and Rabi splitting phenomenology in the
system.
\end{abstract}
\maketitle

The deeply sub-wavelength character of localized surface plasmons
(SPs) provides new avenues for the control of light-matter
interactions at the nanoscale, both in the
weak~\cite{Novotny2011,Giannini2011} and strong coupling
regimes~\cite{Zengin2015,Chikkaraddy2016}. Currently, hybrid
systems comprising metal nanocavities and quantum emitters (QEs)
are attracting much interest not only for their fundamental
implications, but also for their technological prospects in areas
such as photonics~\cite{Ramezani2017} and material
science~\cite{Orgiu2015}. Lately, experimental reports have shown
that the strong light confinement enabled by SPs can unveil
features of microscopic light sources that remain hidden to
propagating fields, such as mesoscopic effects in the electronic
wavefunctions of quantum dots~\cite{Andersen2010} or the
fingerprint of individual chemical bonds in Raman
molecules~\cite{Benz2017}. These advances indicate that in order
to fully seize the potential of QE-SP devices their theoretical
description~\cite{Torma2015} must combine the framework of
macroscopic quantum electrodynamics, accounting for the lossy and
open nature of SP quanta~\cite{Dung1998}, and refined models for
QEs, including ingredients such as rovibrational~\cite{Galego2015}
or polarization degrees of freedom~\cite{Lodahl2017}.

In this Letter, we investigate the impact that light-forbidden
exciton transitions have in QE-SP interactions at the single
emitter level. We consider a V-type three-level system with one
dipolar and one quadrupolar (dipole-inactive) excited states. The
latter are long-lived excitations which present radiative decay
rates typically 5 orders of magnitude lower than dipolar
ones~\cite{Klessingerbook,Rivera2017}. They are effectively
decoupled from propagating light, but recent theoretical
predictions~\cite{Rivera2017,Zurita2002,Filter2012,Alabastri2016}
suggest that the large evanescent field gradients associated to
SPs may allow Purcell enhancing these transitions up to time
scales comparable to light-allowed ones. We explore the influence
of this phenomenon in QE-SP coupling and the formation of
plasmon-exciton-polaritons (PEPs) in an archetypal
nanoparticle-on-a-mirror (NPoM) cavity~\cite{Chikkaraddy2016}.
Using transformation optics~\cite{Pendry2012}, we describe in a
fully analytical manner the near- and far-field characteristics of
the SP modes supported by this structure. This provides deep
physical insights into the population dynamics and the scattering
spectrum of the hybrid QE-SP system, and allows us to reveal the
conditions in which light-forbidden excitons yield a strong
modification of the Purcell enhancement and Rabi splitting
phenomena.

\begin{figure}[!h]
\includegraphics[width=0.93\linewidth]{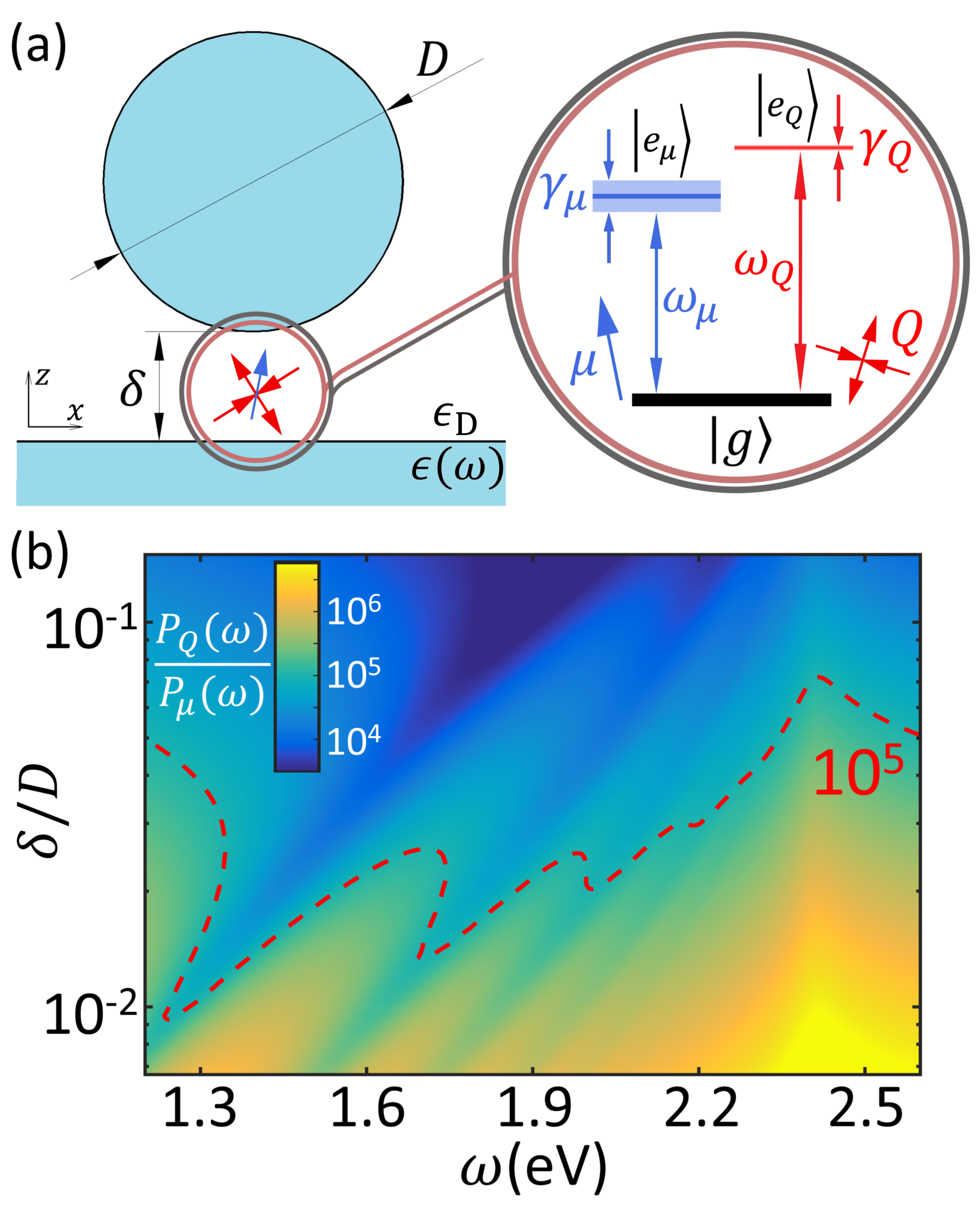}
\vspace{-0.35 cm} \caption{(a) Sketch of the system under study: A
QE is placed at the gap of a metallic NPoM cavity with
$\delta=0.9$ nm and $D=30$ nm. The QE is modelled as a three-level
system with one dipolar ($\mu$) one quadrupolar ($Q$) excitonic
transitions. (b) Ratio between the quadrupolar and dipolar Purcell
factors evaluated at the gap center as a function of the exciton
frequency and gap size.} \label{fig:1}
\end{figure}

Figure~\ref{fig:1}(a) sketches a NPoM cavity with diameter
$D=30$~nm and gap size $\delta=0.9$~nm. The metal permittivity
$\epsilon(\omega)=\epsilon_\infty-\omega_{\rm
p}^2/\omega(\omega-i\gamma)$ is a low-frequency Drude-fitting to
Ag, see Supplemental Material (SM), and the whole structure is
embedded in $\epsilon_{\rm D}=4$. A point-like three-level system
is placed at ${\bf{r}_{\rm QE}=x_{\rm QE}\bf{\hat{x}}+z_{\rm
QE}\bf{\hat{z}}}$ (the coordinate origin is located at the bottom
of the gap). The light-allowed transition has a dipole moment
$\boldsymbol{\mu}=\mu \bf{\hat{z}}$, a natural frequency
$\omega_{\mu}$, and radiative decay rate in vacuum
${\gamma_{\mu}=\gamma_{\mu}(\omega_{\mu})=\omega_{\mu}^3
\mu^2/3\pi\epsilon_0 \hbar c^3}$. The light-forbidden one has a
quadrupole moment
$\boldsymbol{Q}=\tfrac{Q}{\sqrt{2}}[\bf{\hat{z}\hat{x}}+\bf{\hat{x}\hat{z}}]$,
frequency $\omega_{Q}$ and decay rate
$\gamma_{Q}=\gamma_{Q}(\omega_Q)=\omega_Q^5 Q^2/360\pi\epsilon_0
\hbar c^5$. Nonradiative decay in the QE is neglected, and the
orientation of both excitonic moments is chosen to maximize SP
coupling. We take $\mu=0.56$~e$\cdot$nm, $\omega_{\mu}=1.55$~eV
and $\gamma_{\mu}=0.6$~$\mu$eV. These parameters and the NPoM
dimensions are in accordance with the experimental set-up in
Ref.~\cite{Chikkaraddy2016}.

Using transformation optics, we can calculate the scattering Green
function, $\bf{G}(\bf{r},\bf{r}_{\rm QE})$, for the cavity in
Figure~\ref{fig:1}(a)~\cite{Li2016}. In order to keep our
calculations fully analytical for all system configurations, we
model the near-field characteristics of the cavity geometry
through its 2D counterpart~\cite{Aubry2011} (which assumes
translational symmetry along the $y$-direction). Note that recent
reports indicate that the phenomenology of QE-SP strong coupling
in 2D and 3D NPoMs is remarkaby similar~\cite{Demetriadou2017}.
Thus, the total Purcell factors for dipolar and quadrupolar
transitions read~\cite{Raabbook}
\begin{equation}
\begin{split}
P_{\mu}(\omega)&=\tfrac{8}{\mu^2}\rm{Im}\{ \boldsymbol{\mu}
\bf{G}(\bf{r},\bf{r}_{\rm QE}) \boldsymbol{\mu}
\}_{\bf{r}=\bf{r}_{\rm
QE}}, \\
P_{Q}(\omega)&=\tfrac{16 c^2}{\omega^2 Q^2}\rm{Im}\{(\bf{Q}\nabla)
(\nabla^{'} \bf{G}(\bf{r},\bf{r}'))
\bf{Q}\}_{\bf{r},\bf{r}'=\bf{r}_{\rm QE}}. \label{eqPurcell}
\end{split}
\end{equation}

Figure~\ref{fig:1}(b) renders $P_{Q}/P_{\mu}$ versus frequency and
gap size evaluated at the gap center. The red dashed line plots
the contour $P_{Q}/P_{\mu}=10^5$, and sets the parameter region
for which the time scales for light-forbidden and light-allowed
exciton dynamics become comparable. For $\delta/D=0.03$, there
exist two spectral windows fulfilling this condition: below the
lowest-frequency, bright SP mode ($\omega_1$), and at the dark
plasmonic pseudomode ($\omega_{\rm PS}\simeq\omega_{\rm
p}/\sqrt{\epsilon_\infty+\epsilon_{\rm D}}$), which emerges from
the spectral overlapping of high-frequency, tightly confined
SPs~\cite{Li2016}.

\begin{figure}[!t]
\hspace{-0.3 cm}
\includegraphics[width=0.95\linewidth]{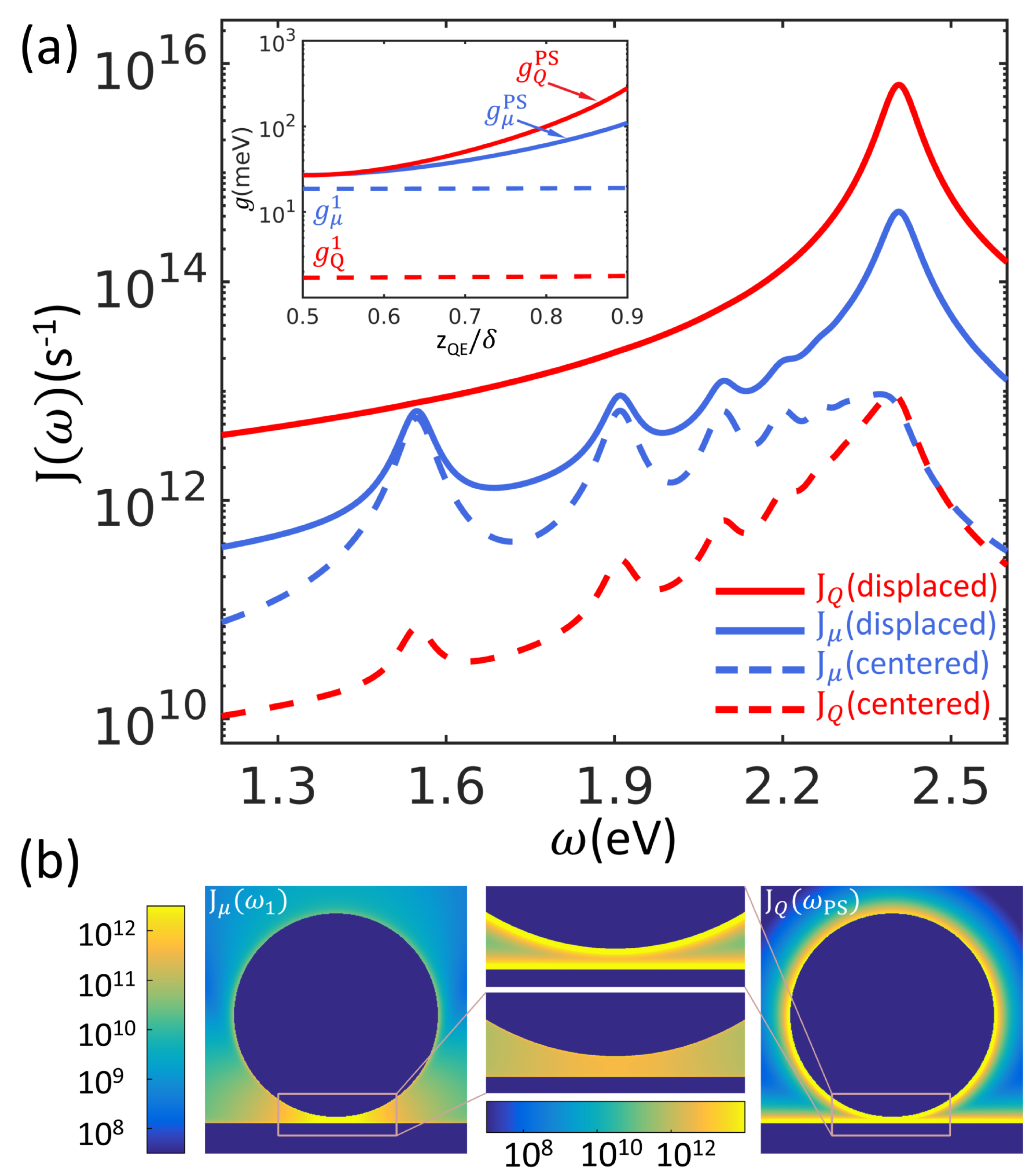}
\caption{(a) Spectral densities ($x_{\rm QE}=0$, $\mu=0.56$
e$\cdot$nm, $Q=0.52$ e$\cdot$nm$^2$) evaluated at $z_{\rm
QE}=\delta/2$ (dashed line) and $7\delta/8$ (solid line). Inset:
Coupling strengths for the lowest-frequency SP (dashed) and the
pseudomode (solid) as a function of $z_{\rm QE}$. (b) Spectral
density maps for the dipolar (left) and quadrupolar (right)
transitions at $\omega_1$ and $\omega_{\rm PS}$, respectively.
Central panels zoom at the gap region (note the difference in
color scales).} \label{fig:2}
\end{figure}

Figure~\ref{fig:2} shows the spectral density,
${J_{i}(\omega)=\tfrac{\gamma_{i}(\omega)}{2\pi}P_i(\omega)}$, for
dipolar ($i=\mu$) and quadrupolar ($i=Q$) excitonic transitions
within the cavity in Figure~\ref{fig:1}(a) and
$Q=0.52$~e$\cdot$nm$^2$. This value is chosen so that the spectral
density at the PS is the same for both excitons at $z_{\rm
QE}=\delta/2$, although $J_{Q}(\omega)$ decays much faster with
decreasing frequency. By displacing the emitter position to
$z_{\rm QE}=7\delta/8$, both spectral densities increase. This
enhancement is much higher in $J_{Q}(\omega)$, which is fully
governed by the PS, and whose maximum is 20 times larger than
$J_\mu(\omega_{\rm PS})$. The bottom panels display the spatial
dependence of the dipolar (quadrupolar) spectral density evaluated
at $\omega_1=1.55$ eV ($\omega_{\rm PS}=2.40$ eV).
$J_{\mu}(\omega_1)$ is large within the gap region and is
uniformly distributed across it. On the contrary,
$J_{Q}(\omega_{\rm PS})$ is tightly confined at all metal surfaces
(see zooms in central panels).

Under the high-quality resonator approximation~\cite{Waks2010},
the analytical expressions for the spectral densities can be
reshaped into a sum of lorentzian terms of the form
\begin{equation}
J_{i}(\omega)=\sum_{n=1,\infty}^{\sigma=\pm1}\frac{\left(g_i^{n,\sigma}\right)^2}{\pi}\frac{\gamma_{n,\sigma}/2}{(\omega-\omega_{n,\sigma})^2+(\gamma_{n,\sigma}/2)^2},
\label{eqJ}
\end{equation}
where again $i=\mu,Q$, $n$ is the SP azimuthal order, and $\sigma$
the parity (even/odd) with respect to the gap center. The SM shows
the validity of Equation~\eqref{eqJ} for the spectra in
Figure~\ref{fig:2}(a). It also presents analytical expressions for
the SP frequencies $\omega_{n,\sigma}$ and decay rates
$\gamma_{n,\sigma}=\gamma+\delta_{\sigma 1}\gamma^{\rm rad}_{n}$
(where $\gamma$ is the Drude damping and $\gamma^{\rm rad}_{n}$
are the radiative rates of even SPs), as well as the QE-SP
coupling constants $g_i^{n,\sigma}$. The inset of
Figure~\ref{fig:2}(a) plots the coupling strengths for dipole and
quadrupole excitons and for the lowest-frequency SP (the
superscript $\sigma=1$ is omitted) and for the pseudomode,
$g_{i}^{\rm PS}=\sqrt{\sum_{n\geq7,\sigma}(g_i^{n,\sigma})^2}$.
Whereas the former do not depend on $z_{\rm QE}$, the latter grow
exponentially as the QE approaches the metal surface, being this
enhancement always larger for the quadrupole exciton.

We use our approach to assess the influence that light-forbidden
transitions have in the exciton population dynamics. We assume
that initially only the light-allowed state is populated
($n_{\mu}(0)=|c_{\mu}(0)|^2=1$ and $n_Q(0)=|c_{Q}(0)|^2=0$) and
analyze how the excited state populations evolve in time. The
Wigner-Weisskopf problem for our system consists in two coupled
integro-differential equations of the form
\begin{equation}
\begin{aligned}
\dot{c}_i(t)=& -\int_{0}^{t} d\tau c_i(\tau) \int_0^\infty
J_{i}(\omega) e^{i(\omega - \omega_i)(\tau-t)} d\omega - \\
&-\int_{0}^{t} d\tau c_j(\tau) \int_0^\infty J_{V}(\omega)
e^{i[(\omega - \omega_j)\tau-(\omega - \omega_i)t]} d\omega,
\end{aligned}\label{eqWW}
\end{equation}
where $i,j=\mu,Q$ ($i\neq j$), and $J_{V}(\omega)$ is given by
Equation~\eqref{eqJ} with
$g_V^{n,\sigma}=\sqrt{g_\mu^{n,\sigma}g_Q^{n,\sigma}}$. This
spectral density feeds the second term in Equations~\eqref{eqWW},
which couples the excitonic populations through the full plasmonic
spectrum supported by the NPoM.

\begin{figure}[!t]
\includegraphics[width=0.95\linewidth]{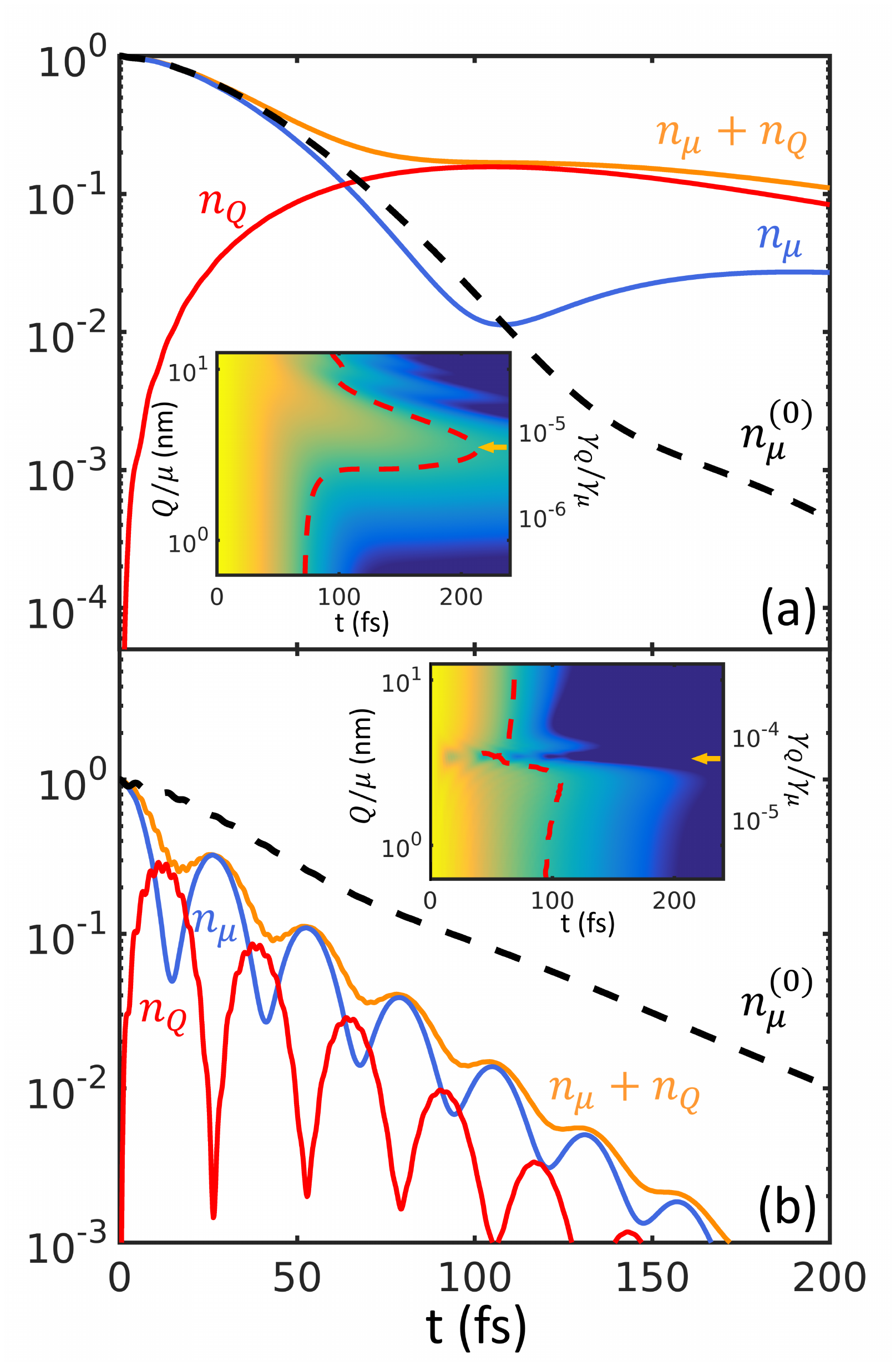}
\vspace{-0.2 cm} \caption{Temporal evolution of the dipolar,
$n_{\mu}$, quadrupolar, $n_{Q}$, and total, $n_{\mu}+n_{Q}$,
excitonic populations with initial condition $n_{\mu}(0)=1$. For
comparison, the dipole population in absence of the
light-forbidden transition, $n_{\mu}^{(0)}$, is also shown. In
panel (a), $\omega_{\mu}=\omega_{Q}=\omega_{1}$,
$\gamma_Q/\gamma_\mu=6\cdot10^{-6}$ and $z_{\rm QE}=\delta/2$. In
panel (b), $\omega_{\mu}=\omega_{1}$, $\omega_{Q}=\omega_{\rm
PS}$, $\gamma_Q/\gamma_\mu=5\cdot10^{-5}$ and and $z_{\rm
QE}=7\delta/8$. Insets: total QE population dynamics as a function
of $\gamma_Q/\gamma_\mu$. Red dashed lines plot the contour
$n_{\mu}+n_{Q}=0.1$. Horizontal arrows indicate the cases
considered in the main panels.} \label{fig:3}
\end{figure}

We investigate the population dynamics in two different
configurations, already introduced in Figure~\ref{fig:2}. In
Figure~\ref{fig:3}(a), the QE is placed at the gap center and the
dipole and quadrupole transition frequencies are at resonance with
the lowest SP mode, $\omega_Q=\omega_\mu=\omega_1$. In
Figure~\ref{fig:3}(b), the QE is in the vicinity of the
nanoparticle surface, $z_{\rm QE}=7\delta/8$, and the quadrupole
transition is shifted to the pseudomode frequency,
$\omega_Q=\omega_{\rm PS}$. In the main panels, the population of
the light-allowed, $n_\mu=|c_\mu|^2$ (blue), and light-forbidden,
$n_Q=|c_Q|^2$ (red), states are shown, as well as the total QE
population, $n_\mu+n_Q$ (orange). They are evaluated at
$Q/\mu=3.5$ nm, which corresponds to
$\gamma_Q/\gamma_\mu=6\cdot10^{-6}$ and
$\gamma_Q/\gamma_\mu=5\cdot10^{-5}$ in panels (a) and (b),
respectively. For comparison, the dipole state populations in
absence of the quadrupole exciton, $n_\mu^{(0)}$, are shown in
black dashed lines. Note that both show a rather monotonic decay
decorated by very shallow oscillations, which can be linked to the
onset of the QE-SP strong coupling regime.

Figure~\ref{fig:3}(a) shows $n_Q$ growing initially, up to
crossing $n_\mu$. At longer times, the quadrupole exciton feeds
population back into the dipole state, and induces a decay in
$n_\mu+n_Q$ which is significantly slower than $n_\mu^{(0)}$. The
fingerprint of the quadrupole exciton is even more remarkable in
Figure~\ref{fig:3}(b), where the fast Rabi oscillations in $n_Q$,
which originate from its strong coupling to the PS, are
transferred to $n_\mu$ as well. The resulting oscillating
$n_\mu+n_Q$ profile decays much faster than $n_\mu^{(0)}$. Thus,
we can conclude that depending on the configuration,
light-forbidden transitions can effectively reduce or enlarge the
QE lifetime, altering significantly the phenomenology of the
Purcell effect in the system. The insets display $n_\mu+n_Q$
versus $\gamma_Q/\gamma_\mu$ and time, revealing that such strong
modifications of QE lifetime due to light-forbidden excitons only
take place within a certain range of $Q$-values in both
configurations.

In order to evaluate the impact that the quadrupole exciton has in
the performance of the QE-SP system as a photonic device, we study
next its far-field scattering spectrum. We model a dark-field
spectroscopy set-up~\cite{Chikkaraddy2016} in which the NPoM is
illuminated by a grazing laser field, $E_{\rm L}$, with frequency
$\omega_{\rm L}$ and polarized along $z$-direction. Due to its
inherent open and lossy nature, describing the scattering
properties of the hybrid system would require, in principle, the
computation of its steady-state density matrix out of a Liovillian
formulation of the problem. However, in the limit of low pumping
($E_{\rm L}\rightarrow0$), we can use a non-hermitian
hamiltonian~\cite{Visser1995} of the form
\begin{eqnarray}
\hat{H} & = & \sum_{n,\sigma}
\tilde{\omega}_{n,\sigma}\hat{a}_{n,\sigma}^{\dagger}\hat{a}_{n,\sigma}
+ \tilde{\omega}_\mu
\hat{\sigma}_{\mu}^{\dagger}\hat{\sigma}_{\mu}+ \omega_Q
\hat{\sigma}_{Q}^{\dagger}\hat{\sigma}_{Q}+
\nonumber \\
& & + \Big(\sum_{n,\sigma,j}
g_{j}^{n,\sigma}\hat{a}_{n,\sigma}^{\dagger}\hat{\sigma}_{j}+E_{\rm
L}e^{-i\omega_{\rm L}t}\hat{M}^\dagger + {\rm h.c.}\Big),
 \label{eqH}
\end{eqnarray}
where $\hat{a}_{n,\sigma}^\dagger$ ($\hat{a}_{n,\sigma}$) and
$\hat{\sigma}^\dagger_{j}$ ($\hat{\sigma}_{j}$) are the creation
(annihilation) operators for SP and QE excitations, and
$\tilde{\omega}_{n,\sigma}=\omega_{n,\sigma}-i\tfrac{\gamma_{n,\sigma}}{2}$
and $\tilde{\omega}_{\mu}=\omega_{\mu}-i\sqrt{\epsilon_{\rm
D}}\tfrac{\gamma_\mu}{2}$ are the complex frequencies for the SPs
and the QE dipole transition, respectively (note that the latter
radiates in a dielectric background $\epsilon_{\rm D}$).

The Hamiltonian in Equation~\eqref{eqH} is equivalent to the one
behind Equations~\eqref{eqJ} except for the last term. This
accounts for the coherent pumping of the system, with
$\hat{M}=\sum_{n}\mu_n\hat{a}_{n,1}+\mu\hat{\sigma}_{\mu}$. Note
that the SP dipole momenta can be obtained (see SM) from the
radiative decay rates as
$\mu_n=\tfrac{\epsilon(\omega_{n,1})-\epsilon_{\rm
D}}{2\epsilon(\omega_{n,1})}\sqrt{3\pi\epsilon_0 \hbar\gamma^{\rm
rad}_{n}c^3/\sqrt{\epsilon_{\rm D}}\omega_{n,1}^3}$. We solve the
Schr\"odinger equation for $\hat{H}$ transformed into the rotating
frame. Using perturbation theory on the parameter $E_{\rm
L}$~\cite{SaezBlazquez2017}, we obtain the steady-state
wavefunction for the QE-SP system, $|\psi_{\rm SS}\rangle$, and
compute its cross section as $\sigma_{\rm sca}=\langle\psi_{\rm
SS}|\hat{M}^{\dagger}\hat{M}|\psi_{\rm SS}\rangle$ (note that
taking advantage of the deeply subwavelength dimensions of the
system, we can drop the near-to-far-field Green's function in our
calculation of the scattering spectrum).

\begin{figure}[!t]
\hspace*{-0.5 cm}
\includegraphics[angle=0,width=1.1\linewidth]{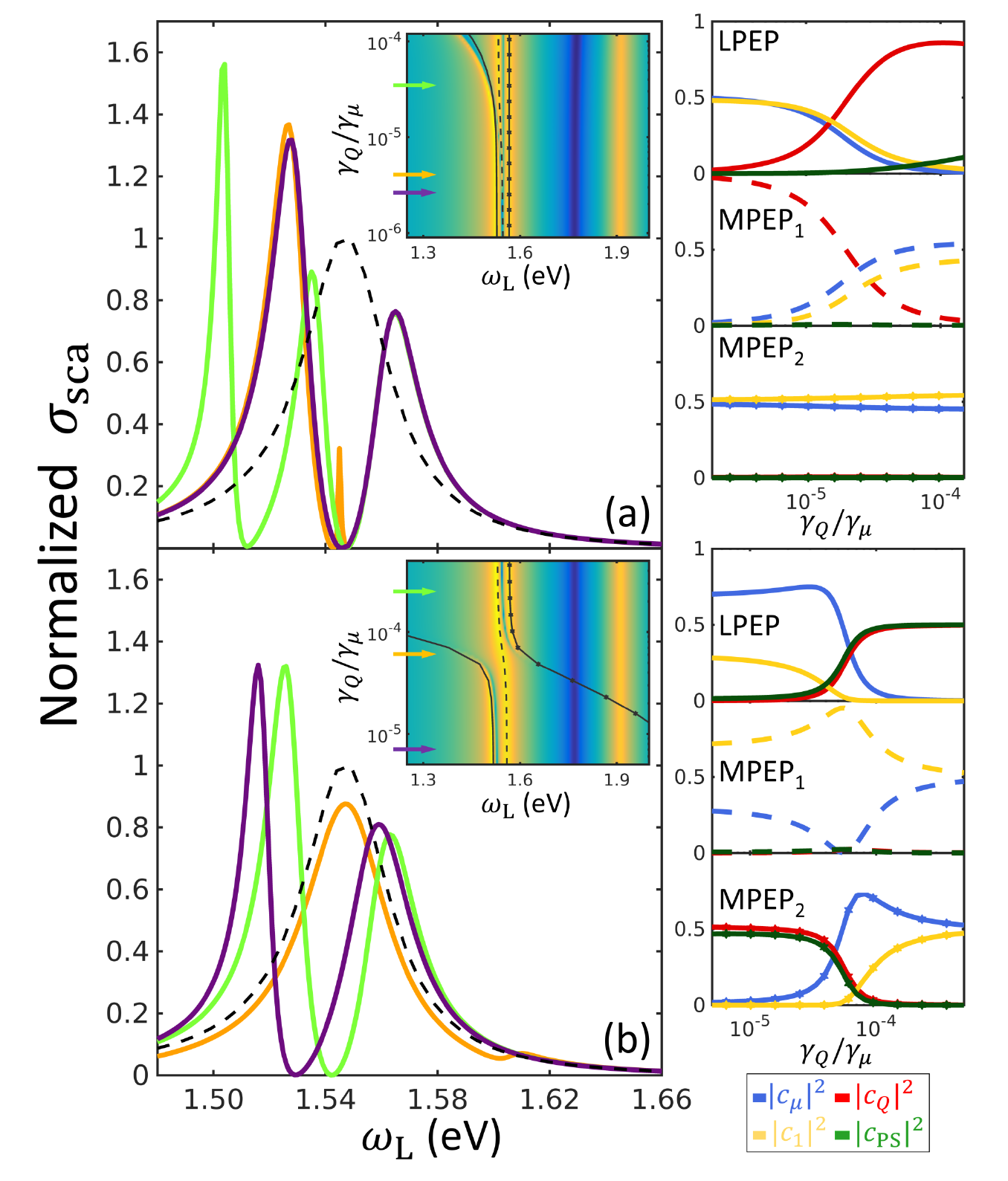}
\vspace{-0.4 cm} \caption{Scattering spectra at the lowest plasmon
resonance for the hybrid the QE-NPoM systems in Figure 3:
$\omega_{\mu}=\omega_{Q}=\omega_{1}=1.55$ eV (a) and
$\omega_{\mu}=\omega_{1}=1.55$ eV, $\omega_{Q}=\omega_{\rm
PS}=2.40$ eV (b). Black dashed lines render $\sigma_{\rm abs}$ for
the bare NPoM, and color solid lines correspond to hybrid systems
with different quadrupole momenta. The insets show the cross
section as a function of the laser frequency and
$\gamma_Q/\gamma_\mu$ (arrows indicate the configurations in the
main panels). Solid, dashed and connected-dotted lines render the
PEP dispersion bands. Right panels plot the Hopfield coefficients
for all cases as a function of $\gamma_Q/\gamma_\mu$.}
\label{fig:4}
\end{figure}

Figure~\ref{fig:4} renders the scattering cross section for the
two configurations considered in Figure~\ref{fig:3}. The spectral
window is centered around $\omega_1$, the frequency of the lowest
SP supported by the NPoM and whose contribution to the far-field
signal is the largest. In both panels, the bare SP spectra is
plotted in black dashed line (see SM for a comparison against
numerical solutions of Maxwell's Equations). Three different
$\gamma_Q/\gamma_\mu$ ratios are considered: orange lines
correspond to the $Q$-values in the main panels of
Figure~\ref{fig:3}, whereas violet and green lines plot spectra
for lower and higher quadrupole momenta, respectively. The insets
display $\sigma_{\rm sca}$ versus laser frequency and
$\gamma_Q/\gamma_\mu$ within a wider spectral frequency, ranging
beyond the scattering peak at $\omega_{2,1}$. Note the presence of
an invisibility dip~\cite{Aubry2011} at 1.77 eV, which originates
from superposition effects among the emission from different SPs.

The violet spectrum in Figure~\ref{fig:4}(a) shows the Rabi
splitting of the bare NPoM scattering peak, an indication of the
formation of PEPs in the system due to the strong coupling between
the lowest SP and the QE dipolar exciton~\cite{Chikkaraddy2016}.
The fingerprint of the quadrupole exciton becomes apparent only at
larger $\gamma_Q/\gamma_\mu$, giving rise to a third peak in
$\sigma_{\rm sca}$ at $\omega_{\rm L}=\omega_1$ (orange line).
This maximum grows and broadens, and its position red-shifts, as
the quadrupole moment increases further (green line). The right
panels in Figure~\ref{fig:4}(a) plot the Hopfield coefficients for
the three PEPs behind these far-field signatures. They are
obtained from the projection of $|\psi_{\rm SS}\rangle$ in the
bare exciton-plasmon basis (for simplicity, we have restricted the
plasmonic Hilbert space to the lowest SP and the PS). Following an
increasing frequency order, the PEPs are labelled as lower, middle
1 and middle 2. A fourth (upper) polariton emerges at $\omega_{\rm
PS}$, not shown here. These panels show that the peak at
$\omega_1$ in Figure~\ref{fig:4}(a) develops when MPEP1 losses
partially its light-forbidden excitonic character. This is in turn
transferred to LPEP, whose associated peak becomes narrower. On
the contrary, MPEP2 is barely affected in this process.
Importantly, this modification in the usual Rabi splitting
phenomenology takes place within the parametric region in which
the Purcell effect in Figure~\ref{fig:3}(a) is reduced and the QE
lifetime is longest.

Figure~\ref{fig:4}(b) reveals that the Rabi splitting profile is
altered even at the lowest $\gamma_Q/\gamma_\mu$ in the second
QE-SP configuration. This higher sensitivity to the quadrupole
exciton is a consequence of the large enhancement that $g_{Q}^{\rm
PS}$ experiences as the QE is displaced across the cavity gap, see
Figure~\ref{fig:2}(a). Note that due to the large spectral
detuning between $\omega_1$ and $\omega_{\rm PS}$, the scattering
dip is no longer at $\omega_{\rm L}=\omega_1$. Remarkably, by
increasing the quadrupole moment, the lowest frequency peak
vanishes, and the spectrum for the hybrid system resembles very
much the one of the bare NPoM (orange line). This profile can be
observed only within a narrow range of $Q$-values, beyond which a
symmetric Rabi splitting spectrum, very similar to the one
observed at $Q=0$ is recovered (green line). The contourplot in
the inset and the Hopfield coefficients shed light into this
evolution of $\sigma_{\rm sca}$. In this case, MPEP2, which
initially originates from the strong coupling between the
quadrupole exciton and the PS, becomes strongly Rabi-shifted
towards $\omega_1$. For large enough $Q$, anti-crossing among PEP
bands takes place. As a result, LPEP and MPEP2 lose their content
on the lowest SP mode, which in turn is transferred to MPEP1. In
these conditions, light-allowed and light-forbidden QE states
interact only through the PS and become completely dark. The Rabi
splitting in $\sigma_{\rm sca}$ vanishes but, as shown in
Figure~\ref{fig:3}(b), the large excitonic couplings to the PS
lead to a fast oscillations in the QE population and an effective
shortening of its lifetime.

To conclude, we have investigated the influence of light-forbidden
excitons in plasmon-emitter coupling at the gap of
nanoparticle-on-a-mirror cavity. We have developed a fully
analytical description of the spectral densities for both dipolar
and quadrupolar (dipole-inactive) transitions, obtaining the
plasmonic natural frequencies and decay rates, as well as the
different sets of plasmon-exciton coupling strengths. Using our
approach, we have explored the near-field population dynamics and
far-field scattering spectrum of these hybrid systems, revealing
the conditions in which the quadrupole excitons lead to strong
modifications in the Purcell effect and Rabi splitting
phenomenology. Our results prove that, by means of surface
plasmons, the internal degrees of freedom of microscopic light
sources can be exploited for the realization of richer and more
versatile platforms for polaritonic applications.

The authors thank Elena del Valle and Alejandro Gonz\'alez-Tudela
for fruitful discussions. This work has been funded by the Spanish
MINECO under contracts FIS2015-64951-R, MDM-2014-0377-16-4 and
through the ``Mar\'ia de Maeztu'' programme for Units of
Excellence in R\&D (MDM-2014-0377), as well as the EU Seventh
Framework Programme under Grant Agreement
FP7-PEOPLE-2013-CIG-630996.

\end{document}